\documentclass[aps,prl,reprint,superscriptaddress,amsmath,a4paper]{revtex4-1}
\usepackage{hyperref}
\usepackage[dvips]{graphicx}

\newcommand{\txt}[1]{\mathrm{#1}}
\newcommand{\kB}{k_\txt{B}}

\newcommand{\Ec}{E_\txt{c}}
\newcommand{\RT}{R_\mathrm{T}}
\newcommand{\RK}{R_\txt{K}}
\newcommand{\CSigma}{C_\Sigma}
\newcommand{\unit}[1]{\ \mathrm{#1}}
\newcommand{\etal}{\emph{et al.}}

\begin{document}
\title{Test of Jarzynski and Crooks fluctuation relations in an electronic system}
\author{O.-P. Saira}
\affiliation{Low Temperature Laboratory, Aalto University, P.O. Box 15100, FI-00076 AALTO, Finland}
\affiliation{Department of Applied Physics/COMP, AALTO University, P.O. Box 14100, FI-00076 AALTO, Finland}
\author{Y. Yoon}
\affiliation{Low Temperature Laboratory, Aalto University, P.O. Box 15100, FI-00076 AALTO, Finland}
\author{T. Tanttu}
\affiliation{Department of Applied Physics/COMP, AALTO University, P.O. Box 14100, FI-00076 AALTO, Finland}
\author{M. M\"ott\"onen}
\affiliation{Department of Applied Physics/COMP, AALTO University, P.O. Box 14100, FI-00076 AALTO, Finland}
\affiliation{Low Temperature Laboratory, Aalto University, P.O. Box 15100, FI-00076 AALTO, Finland}
\author{D. V. Averin}
\affiliation{Department of Physics and Astronomy, Stony Brook University, SUNY, Stony Brook, NY 11794-3800, USA}
\author{J. P. Pekola}
\affiliation{Low Temperature Laboratory, Aalto University, P.O. Box 15100, FI-00076 AALTO, Finland}

\begin{abstract}
Recent progress on micro- and nanometer scale manipulation has opened the possibility to probe systems small enough that thermal fluctuations of energy and coordinate variables can be significant compared with their mean behavior. We present an experimental study of nonequilibrium thermodynamics in a classical two-state system, namely a metallic single-electron box. We have measured with high statistical accuracy the distribution of dissipated energy as single electrons are transferred between the box electrodes. The obtained distributions obey Jarzynski and Crooks fluctuation relations. A comprehensive microscopic theory exists for the system, enabling the experimental distributions to be reproduced without fitting parameters. 
\end{abstract}
\maketitle

%individual complex biomolecules~\cite{Liphardt2002,Collin2005,Junier2009} and colloidal particles~\cite{Wang2002}
%single-electron box~\cite{Averin1985,Buttiker1987,Lafarge1991}

Everyday concepts such as heat and mechanical work have their foundation in dynamics at the atomic and molecular scale, which is typically unobservable in macroscopic samples. Advances in the synthesis and manipulation of biological matter have made it possible to study the response of individual molecules~\cite{Liphardt2002,*Collin2005,*Junier2009} to mechanical forces, also see Ref.~\cite{Alemany2011} for a recent review. In these experiments, the external force is modulated following a predetermined protocol and the resulting work $W$ exerted upon the system is measured. Consequently, the fluctuating nature of the thermodynamic quantity `work' is revealed, as repeated measurements yield different results depending on the microscopic trajectory traversed by the molecule. Similar studies have been performed on colloidal particles~\cite{Wang2002,*Blickle2006} and mechanical oscillators~\cite{Douarche2005}. Fluctuation theorems such as the Jarzynski equality~\cite{Jarzynski1997} enable one to infer path-independent equilibrium free energy differences from the statistics of irreversible measurement protocols. The Jarzynski equality (JE) states that
\begin{equation}
\left< e^{-W / \kB T} \right> = e^{-\Delta F / \kB T}, \label{eq:JE}
\end{equation}
where the ensemble average is taken over repetitions of the force protocol starting from an equilibrium state at temperature $T$, and $\Delta F$ is the difference in free energy between the final and initial values of the external control parameters. As we demonstrate experimentally, this equation applies also to a single-electron device operated in a dilution refrigerator at temperatures of about 200~mK. An electrostatic gate drive plays the role of the time-dependent mechanical force that has been missing in previous studies of fluctuation theorems in electronic systems~\cite{Garnier2005,*Utsumi2010,*Kung2012}, and is crucial for studying JE~(\ref{eq:JE}) and Crooks fluctuation relations discussed below.

\begin{figure}[h]
\includegraphics[width=.49\textwidth]{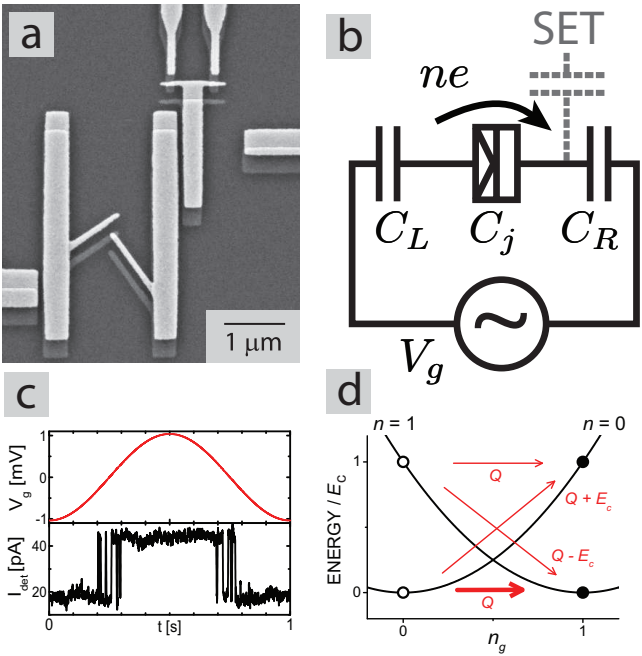}
    \caption{
(a) Scanning electron micrograph of the active area of the measured sample which shows metallic films fabricated on an oxidized silicon wafer by e-beam lithography and shadow evaporation technique~\cite{FultonDolan}. Two shifted copies of the original resist mask pattern lie on the surface: Copper layer appears brighter compared to oxidized aluminum.  Tunnel junctions are formed in the overlap regions between the two films. The single-electron box is located on the left, and the SET electrometer at the top. The tips of two gate electrodes that are used to control the electrostatics of the box and the electrometer are visible at the left and right edges. (b) Simplified circuit diagram of the system. The galvanically isolated single-electron box is connected capacitively to its environment via $C_L$ and $C_R$, and to the electrometer as illustrated by the dashed gray line.  (c) Full period of the sinusoidal drive signal (top) applied to the control gate, and one instance of electrometer response (bottom). The drive frequency is 1 Hz, and the amplitude is equal to one gate modulation period of the box. (d) Energy level diagram of the system for the two lowest-energy charge states. Black parabolas represent the charging energy of the system in the states $n = 0$ and $n = 1$ as a function of the extrnally controlled gate charge $n_g$. Possible starting and ending configurations are indicated by open and filled circles, respectively, and the red annotation text indicates the dissipated work $W - \Delta F$ for each trajcetory type, illustrating that work separates into dissipated heat $Q$ and change of internal energy.
}
\end{figure}

The system that is the subject of our study is a single-electron box~\cite{Averin1985,*Buttiker1987,*Lafarge1991} as depicted in the scanning electron micrograph of Fig.~1(a). The box is formed by two metallic electrodes that are electrically connected through a tunnel junction with a small electric capacitance $C_j$. The complete circuit diagram and naming of capacitors is shown in Fig.~1(b).
%For the purposes of this work, the two electrodes can be considered ideal capacitor plates where surface charges can move about without any significant dissipation.
%Tunnel resistance $\RT$ of the junction is large compared to the resistance quantum $\RK$, so that charge transport through the junction is described by the basic theory of single-electron tunneling~\cite{AverinLikharev,IngoldNazarov}.
The sole degree of freedom that is of interest here is the number $n$ of electrons transferred from the left to the right electrode. %In writing down the system electrostatics, we ignore back-action from the weakly coupled charge detector, but account for the work performed on the tunneling electrons by the voltage source maintaining the gate voltage $V_g$ of the control gate.
The relevant $n$-dependent part of the electrostatic energy of the single-electron box is given by~\cite{PekolaNote2012}
\begin{equation}
U = \Ec (n^2- 2n n_g),\label{eq:H}
\end{equation}
where $\Ec = e^2/(2 \CSigma)$ is the characteristic unit of charging energy of the box, $e$ is electron charge, $\CSigma = C_j + C_g$ is the total capacitance of the box, $C_g = \left(C_L^{-1} + C_R^{-1}\right)^{-1}$ is the effective gate capacitance, and $n_g = C_g V_g/e$ is the gate charge in units of $e$. The charge number $n$ changes by $\pm 1$ in the process of electron tunneling across the junction. In the temperature range of strong Coulomb blockade $T \ll \Ec / \kB$, thermal excitations of $n$ are exponentially unlikely when the gate charge $n_g$ in Eq.~(2) is an integer~\cite{Lafarge1991}. Thus, the charge number $n$ can be driven between two adjacent charge states, say, $n = 0$ and $n = 1$, by a protocol $n_g(t)$ that ramps the gate charge between the values $n_g=0$ and $n_g=1$. This is demonstrated in Fig.~1(c).

We employ a readout that yields the heat $Q = W - \Delta U$ deposited into the electrodes in such driven transitions. The theoretical analysis presented in Refs.~\cite{Averin_Statistics,PekolaNote2012} establishes that JE can be written in terms of $Q$ as
\begin{equation}
\left< e^{-Q / \kB T} \right> = 1,\label{eq:JE_Q}
\end{equation}
provided that the drive protocol is such that the $n = 0\rightarrow 1$ transition always occurs. This condition is realized in the present experiment to an accuracy of $10^{-4}$~\cite{supmat}. For completeness, we illustrate  in Fig.~1(d) the relationship between $Q$ and the dissipated work $W_\txt{dis} = W - \Delta F$ appearing in the fluctuation relations for other trajectory types as well. One can similarly write the Crooks fluctuation theorem~\cite{Crooks1999} in terms of $Q$ as
\begin{equation}
\frac{P_F(-Q)}{P_R(Q)} = e^{-Q/\kB T}, \label{eq:CFT_Q}
\end{equation}
where $P_F$ and $P_R$ are the probability distributions of $Q$ when the system is driven in forward (F) or reverse (R) directions, respectively.  Importantly, the above formulations of the fluctuation theorems can be applied without detailed knowledge of the internal dynamics of the system, thus retaining their universality and usefulness.

Charge tunneling is governed thermally by the excitations of conduction electrons in the box electrodes that couple to the bath of lattice phonons. Hence, switching dynamics between the different charge states $n$ is dissipative. The amount of energy deposited into the two electrodes in a single tunneling event equals the difference of the chemical potentials of the electrodes at the time of the tunneling, which is essentially instantaneous on the other relevant timescales in the problem~\cite{AverinLikharev,*IngoldNazarov}. The chemical potential difference is given by the change in energy $U$ of Eq.~(\ref{eq:H}) in response to a change $\Delta n=\pm 1$ of the charge number $n$. In general, the $n$-trajectory consists of a random number $N$ of successive back-and-forth tunneling events, and hence the total heat generated in such a trajectory is~\cite{Averin_Statistics}
\begin{equation}
Q = 2\Ec\sum_{k=1}^{N} \pm \left(n_g(\tau_k) - \frac{1}{2}\right),\label{eq:Wdis2}
\end{equation}
where $\tau_k$ is the stochastic time instant of the $k$th tunneling event, and the sign is the same as for $\Delta n$ in the event. Because of the intrinsic randomness of the tunneling events, the heat $Q$ fluctuates from one gate voltage ramp to another. In the experiment, the system was driven with a sinusoidal excitation corresponding to $n_g(t) = \frac{1}{2} - \frac{1}{2} \cos\left(2\pi f t\right)$ with frequencies $f$ ranging from 1 to 20~Hz. These frequencies are sufficiently slow so that the charge state always assumes its minimum energy value at the turning points of the drive, when $n = n_g(t) =$ 0 and 1, respectively. Hence, each half cycle from 0 to 1, and similarly from 1 to 0, can be considered an independent realization of the control protocol. 

We perform the heat readout by detecting the electron tunneling events by a capacitively coupled single-electron transistor (SET)~\cite{supmat}. Equation~(\ref{eq:Wdis2}) yields the heat $Q$ in terms of $\Ec$ for an individual $n_g(t)$ sweep. One can thus utilize the experimental $Q$ distributions in two ways: Using values of $\Ec$ and $T$ determined by independent means, validity of Eqs.~(\ref{eq:JE_Q}) and (\ref{eq:CFT_Q}) can be tested. On the other hand, accepting Eqs.~(\ref{eq:JE_Q}) or (\ref{eq:CFT_Q}), one can determine the ratio $\Ec / \kB T$, and furthermore find $\Ec$ by multiplying this ratio with the independently measured temperature of the sample holder.

%For the parameter values realized in the experiment, the probability of trajectories with %$n(t_\txt{end}) = 0$ is of the order of $10^{-8}$ and their contribution to the Jarzynski %exponential average $\left< \exp\left(-\beta W_\txt{dis}\right) \right>$ is less than $10^{-4}$.

The preceding discussion is independent of the details of the charge tunneling rates in the single-electron box. However, in our case it is possible to analyze the fluctuation relations also from a microscopic point of view. Charge transport through a Cu/AlO$_x$/Al normal metal--insulator--superconductor (NIS) tunnel junctions occurs via thermally activated (TA) $1e$ events described by the orthodox theory~\cite{AverinLikharev,*IngoldNazarov}, provided that (i) the tunneling resistance $\RT$ of the junction is high compared to the resistance quantum $\RK \simeq 25.8\ \mathrm{k\Omega}$, (ii) quasiparticles in the electrodes obey an equilibrium thermal distribution, and (iii) coupling of stray microwaves to the junction has been prevented by appropriate shielding and filtering in the construction of the sample stage and signal lines. Realization of these conditions in NIS single-electron devices, including the back-action from the capacitively coupled elecctrometer, has been studied in detail in recent years \cite{Saira_EAT, Saira_SuperAl}. Based on these studies, non-thermal charge transport is expected to be negligible at least above temperatures of 150~mK. In particular, overheating of the superconducting electrode in the present design is diminished by the fact that quasiparticle excitations can relax to the overlapping normal metal through the oxide barrier. Observed stochastic switching of the system between charge states at a fixed value of gate charge near degeneracy can be directly fitted to the tunneling rates predicted by the orthodox theory~\cite{supmat}. We extract values $\Delta = 218\pm3\unit{\mu e V}$, $\Ec/\kB = 1.94\pm0.05\unit{K}$, and $R_T = 100\pm{13}\unit{M\Omega}$ for the superconducing gap parameter, charging energy of the box, and tunneling resistance of the box junction, respectively. For the detector SET, we obtain $\RT = 0.63\unit{M\Omega}$ and $\Delta = 211\unit{\mu e V}$ from a fit to the measured $I$-$V$ characteristics.

\begin{figure}[ht!]
\includegraphics[width=.49\textwidth]{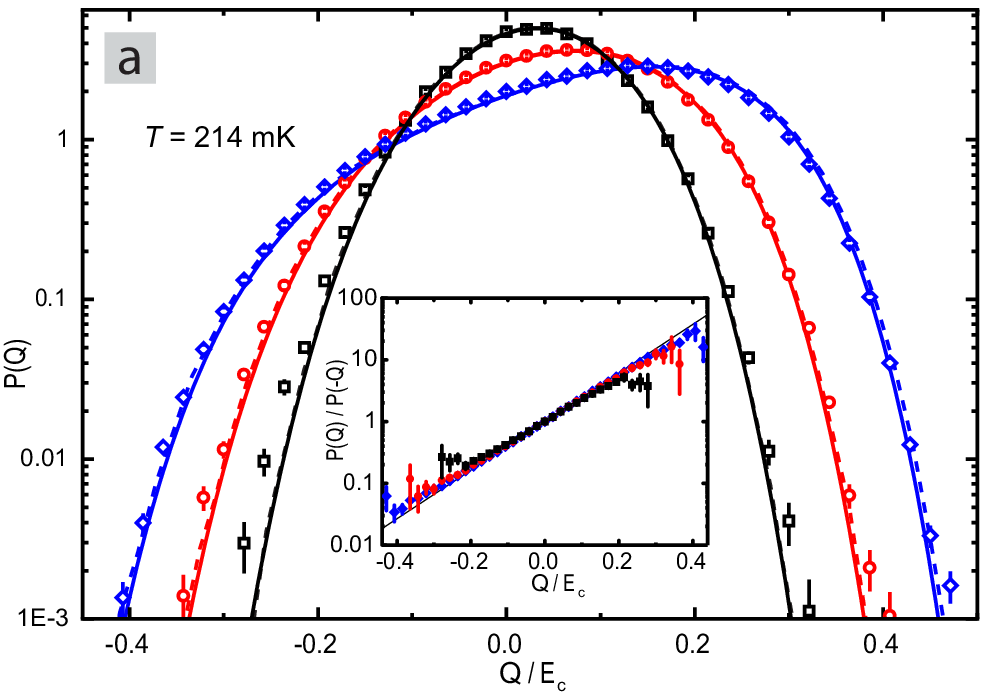}
\includegraphics[width=.49\textwidth]{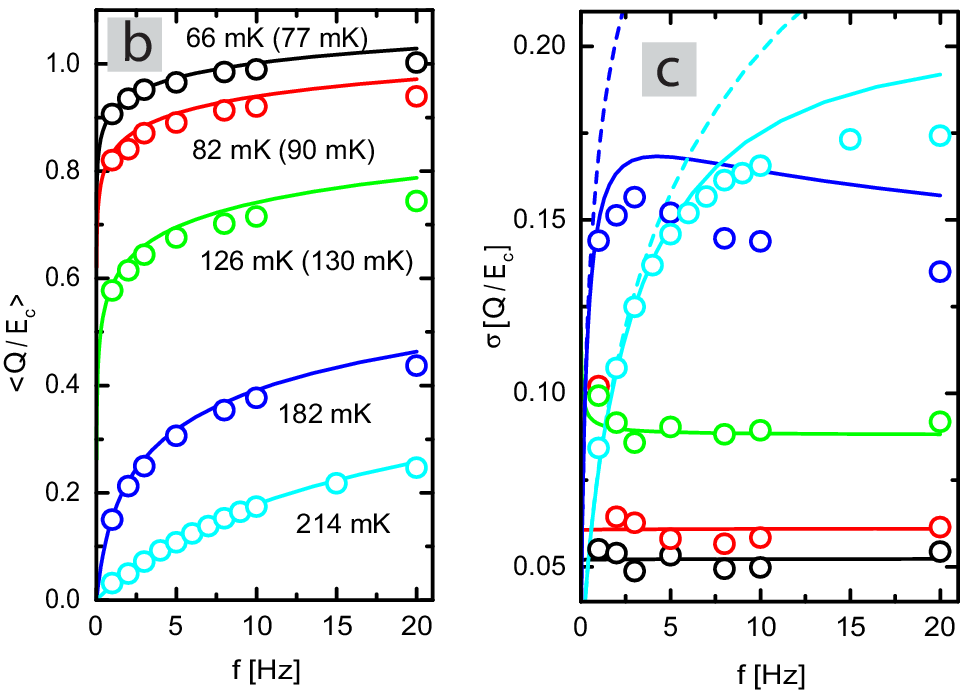}
	\caption{
(a) Measured distribution of the generated heat at drive frequencies 1 Hz (black squares), 2 Hz (red circles), and 4 Hz (blue diamonds). The solid lines are exact theoretical predictions for the independently determined sample parameter values. The dashed lines show the results of Monte Carlo simulations, where the finite bandwidth of the detector was included in the model. Inset: $P(Q) / P(-Q)$ ratio for the experimental distributions. Solid line shows the result from the Crooks fluctuation theorem. (b), (c) First and second moments, respectively, of the $Q$ distribution at different drive frequencies and bath temperatures. Markers are experimental data, and solid lines are exact theoretical predictions as in part (a). For the lowest bath temperatures, the theoretical curves have been calculated using a slightly elevated electron temperature to account for non-ideal thermalization as discussed in the text. The temperature used in the calculation is given in brackets if it differs from the sample stage temperature. In panel (c), dashed lines represent the distribution width inferred from the FDT formula $\left<\left(Q - \left<Q\right>\right)^2 \right> = 2 k T \left<Q\right>$ using the theoretical value of the first moment $\left<Q\right>$.
}
\end{figure}

The experimentally obtained $Q$ distributions for drive frequencies 1, 2 and 4~Hz are presented in Fig.~2(a).
%Acquiring the raw data for each of the experimental distributions took about 18~hours of measurement time.
The distributions were measured at a bath temperature of 214~mK where the thermally activated tunneling rate at degeneracy was 70~Hz, which is well within the detector bandwidth of about 1~kHz. In addition, we have similar data but in smaller quantities for driving frequencies from 5~Hz to 20~Hz. At frequencies higher than this, the observed distributions deviate significantly from the theoretical prediction due to systematic errors arising from finite readout bandwidth and uncertainty in the event timing. On the other hand, driving frequencies lower than 1~Hz make the measurement susceptible to $1/f$ type charge noise that is ubiquitous in metallic single-electron devices~\cite{Starmark1999}. In Fig.~2(a), we show also the exact theoretical distributions based on charge kinetics described by the orthodox theory and sample parameters obtained in the manner described above. The theoretical and experimental distributions are in excellent agreement.

To assess quantitatively the systematic error due to finite detector bandwidth, we show in Fig.~2(a) also the distributions obtained from Monte Carlo simulations that incorporate a finite detector rise-time before threshold detection. Visually, the change in the shape of the distribution functions appears small. Quantitatively, we can assess the accuracy of the readout by evaluating the exponential average $\left<e^{-Q / \kB T} \right>$, which equals 1 for the ideal thermally activated kinetics. From the Monte Carlo simulations, we obtain 1.006 for the 1 Hz and 2 Hz cases, and 1.012 for the 4~Hz case. For the experimental distributions, evaluation of the Jarzynski average yields $1.033 \pm 0.003$ (for 1~Hz drive), $1.032 \pm 0.003$ (2~Hz) and $1.044 \pm 0.004$ (4~Hz), when using the independently determined $\Ec/\kB T$ ratio as described above. The stated uncertainty is the unbiased estimate for the standard deviation of the mean based on the observations, not including the uncertainty of the value of $E_c/\kB T$. As the relative uncertainty of the independent $\Ec$ estimate is 3\%, the Jarzynski equality is shown to hold within experimental accuracy, accounting for the 1\% bias from finite detector bandwidth. Conversely, starting from the assumption that JE holds for the experimental distributions, we obtain an estimate $\Ec / \kB = 1.91 \pm 0.03\unit{K}$.

The possiblity to evaluate numerically the theoretical $Q$ distribution to a high accuracy enables us to assess the magnitude of sampling error in the experiment. From the acquired experimental data, the distribution could be determined for a range of $Q / \Ec$ values where $P(Q) > 10^{-3}$. Evaluating numerically the contribution to JE outside this interval, we see that the deviation due to sampling errors is of the order $10^{-3}$.

Concerning the Crooks fluctuation theorem in the form of Eq.~(\ref{eq:CFT_Q}), the forward and backward distributions coincide in the present case of a gate drive that is antisymmetric with respect to the degeneracy point. Hence, we present the experimental $P(Q)/P(-Q)$ ratio as a function of heat $Q$ in the inset of Fig.~2(a). On a semilogarithmic plot, one expects a linear dependence with the slope equal to $1/ \kB T$, independent of frequency. The experimental data adheres to this quite well, but there is a tendency towards less steep slopes for a lower frequency. This feature can be reproduced in our simulations by assuming a broadening of the $Q$ distributions due to an additive Gaussian noise having a r.m.s. amplitude of $0.0035\,\Ec = 0.032\,\kB T$ independent of driving frequency. We attribute the broadening to the residual background charge noise that induces fluctuations in the exact position of the degeneracy point.

In Figs.~2(b),\,(c) we present the mean generated heat $\mu = \left<Q\right>$ and distribution width $\sigma = \sqrt{\left<(Q - \mu)^2\right>}$ for different bath temperatures in the range $66 - 214\unit{mK}$ and drive frequencies $1-20\unit{Hz}$. The experimental results agree well with the values obtained from numerical simulations performed in the same manner as for Fig.~2(a). Theoretical results for the moments of $Q$ distribution are presented in Ref.~\cite{Averin_Statistics} for a normal state box, and similar results hold for the first two moments in the present NIS case as well. For sufficiently low $f$, the box remains close to local equilibrium during the gate voltage drive. In this case, the transfer of heat $Q$ into the box reservoirs can be viewed as a linear response to small deviations from the equilibrium caused by the drive. As a linear response, this process satisfies the classical fluctuation-dissipation theorem (FDT) which takes the form $\sigma = \sqrt{2\kB T \mu}$. This implies that in the adiabatic limit, the generated heat vanishes not only on average, but for all individual tunneling trajectories. Local equilibrium also implies that the distribution of heat $Q$ is Gaussian~\cite{Averin_Statistics} similarly to all equilibrium thermodynamic fluctuations. The behavior is evident in the experimental data for $T$ = 182~mK and 214~mK, where the slowest drive frequencies produce an almost Gaussian $Q$ distribution, whereas the distributions become strongly non-Gaussian as the frequency is increased. At lower temperatures, the adiabatic threshold frequency is well below 1~Hz and thus inaccessible in the present experiment. For the simulations at the lowest temperatures, we had to use somewhat higher temperatures than those indicated by the sample stage thermometer in order to reproduce the experimental data [see Fig.~2(b)]. At low temperatures, charge kinetics is expected to depart from the basic thermal activation model with a single heat bath, as discussed earlier. Note that our test of the fluctuation relations is not carried out in this regime.

%In conclusion, we have demonstrated a novel readout scheme for heat generated by the driven dynamics of an electronic system, allowing us to study non-Gaussian thermodynamic fluctuations with high fidelity. Charge kinetics of the studied single-electron system are well understood, allowing for a detailed comparison between theory and experiment, and analysis of error sources in the readout.
Thermodynamics of the driven transitions in electronic systems studied in this work will play an important role in the development of reversible information processing devices~\cite{Toyabe2010,*Schaller2011,*Averin2011}. We also envision this work to introduce a fruitful testbed of non-equilibrium fluctuation theorems in qualitatively new settings such as engineered environments. It is possible to realize experimentally a regime where the charge transitions are dominated by coupling to an external nonequilibrium environment~\cite{Saira_EAT}. A fully superconducting box~\cite{Nakamura1999} should enable the study of thermodynamic fluctuations in true quantum regime, which is mostly an unexplored territory in experiments at the moment.

This work has been supported by the Academy of Finland, V\"ais\"al\"a Foundation, ESF research network program EPSD, and European Community's Seventh Framework Programme under Grant Agreement No.~238345 \mbox{(GEOMDISS)}. We thank F.~Ritort, T.~Ala-Nissil\"{a}, and A.~Kutvonen for discussions during composing of the manuscript. The authors have no competing financial interests.

\end{document}

\title{SUPPLEMENTARY MATERIAL to ``Test of Jarzynski and Crooks fluctuation relations in an electronic system''}
\author{O.-P. Saira}
\affiliation{Low Temperature Laboratory, Aalto University, P.O. Box 15100, FI-00076 AALTO, Finland}
\affiliation{Department of Applied Physics/COMP, AALTO University, P.O. Box 14100, FI-00076 AALTO, Finland}
\author{Y. Yoon}
\affiliation{Low Temperature Laboratory, Aalto University, P.O. Box 15100, FI-00076 AALTO, Finland}
\author{T. Tanttu}
\affiliation{Department of Applied Physics/COMP, AALTO University, P.O. Box 14100, FI-00076 AALTO, Finland}
\author{M. M\"ott\"onen}
\affiliation{Department of Applied Physics/COMP, AALTO University, P.O. Box 14100, FI-00076 AALTO, Finland}
\affiliation{Low Temperature Laboratory, Aalto University, P.O. Box 15100, FI-00076 AALTO, Finland}
\author{D. V. Averin}
\affiliation{Department of Physics and Astronomy, Stony Brook University, SUNY, Stony Brook, NY 11794-3800, USA}
\author{J. P. Pekola}
\affiliation{Low Temperature Laboratory, Aalto University, P.O. Box 15100, FI-00076 AALTO, Finland}

%\begin{abstract}
%\end{abstract}
\maketitle

\setcounter{figure}{0}
\renewcommand{\thefigure}{S\arabic{figure}}

We monitor the time-resolved tunneling events by a capacitively coupled SET electrometer as in Refs.~\cite{Fulton1991,Saira_EAT, Saira_SuperAl}. The electrometer is voltage biased and tuned to a charge-sensitive operation point. The current is read out with a room-temperature transimpedance amplifier. Cross-coupling between the gate of the electron box and the SET island is compensated so that the detector current assumes the same value for a fixed charge state $n$ of the box regardless of the control gate voltage. At temperatures $T \lesssim \Ec/\kB$, the detector current will display a characteristic Coulomb staircase as a function of $V_g$. The conversion factor $C_g/e$ from the applied gate voltage $V_g$ to gate charge $n_g$ can be accurately determined from the observed period of the staircase. Slow drifts in background charge during data acquisition were compensated using a first order closed-loop control with a time constant of 100~s.

To apply Eq.~(5) of the main article, we have analyzed the experimental traces with a non-hysteretic threshold detector for determining the switching times $\tau_k$. Use of the threshold detection rejects most of the charge noise in the readout, and the residual noise appears as small jitter in the switching times. Besides intrinsic charge noise, the finite bandwidth of the charge readout leads to two additional error sources. First, there is a constant time delay between the instant of the tunneling event and the time when the filtered detector signal crosses the threshold level. In our setup, most of the time delay and bandwidth limitation originates from the digital low-pass filter employed to reduce the current noise of the readout prior to the threshold detection. The constant time delay has been compensated for in the analysis of the experimental traces. Secondly, finite bandwidth can cause back-and-forth transitions occurring in a rapid succession to be missed altogether~\cite{Naaman_Rates}. To assess the effect of the finite detector bandwidth on the signal processing algorithms, we performed Monte Carlo simulations using a detector step response function extracted from the experimental traces. The results of these simulations are discussed in the main article.

We have measured the rate of thermally activated (TA) transitions as a function of gate charge and bath temperature. Part of that data is presented in Fig.~S1(a). The TA one-electron tunneling rates obey the detailed balance, \ie, $\Gamma(-E) = e^{-E/\kB T} \Gamma(E)$, where $E$ is the chemical potential difference between the electrodes. Hence, $\Ec$ can be determined from the tunneling rates observed at a fixed gate charge $n_g$ by utilizing $\kB T \ln (\Gamma_{0\rightarrow 1}/\Gamma_{1\rightarrow 0}) = 2\Ec(n_g - 1/2)$. We have estimated $\Ec = 1.94\pm0.05\unit{K}$ in this manner from data obtained at different temperatures, see Fig.~S1(b). Furthermore, for $E \ll \Delta$ and $\kB T \ll \Delta$, which is the case in the experiment, one can write the approximation
\begin{equation}
\Gamma(E) = \sqrt{\frac{\pi \kB T \Delta}{2}} \frac{e^{-\Delta/\kB T}}{e^2 \RT} \left[1 + e^{E/\kB T} \right], \label{eq:GammaE}
\end{equation}
where $\Delta$ is the superconducting gap parameter, and $\RT$ is the tunneling resistance. We have determined $\RT$ and $\Delta$ by a nonlinear fit to the tunneling rates observed at different temperatures at degeneracy, \ie, $E = 0$ above. The observed rates are plotted in Fig.~S1(c) together with the rates predicted by Eq.~(1) for the fitted sample parameters.

Finally, for a quantitative assessment of the $W_\txt{dis} = Q$ approximation employed in the interpretation of the data, we have calculated numerically the full $Q$ distributions also for the less likely $0\rightarrow1$, $1\rightarrow1$, and $1\rightarrow0$ trajectory types using the experimental sample and operation parameters stated in the main article. Given that $E_c / \kB = 1.94\unit{K}$, the probability to start (finish) the gate trajectory in a state other than $n = 0$ ($n=1$) is approximately $10^{-4}$ at the highest temperature of 0.214~K for which we present experimental data. Hence, the distribtions for $W_\txt{dis}$ and $Q$ are expected to differ only at the $10^{-4}$ level. The $Q$ distributions can be combined into the full $W_\txt{dis}$ distribution using the relations presented in Ref.~\cite{PekolaNote2012}. The resulting distributions are shown in Fig.~S2, and one observes that the distributions for $W_\txt{dis}$ and $Q_{0\rightarrow1}$ coincide, hence validating the approximation used in the article. 

\begin{figure}
    \includegraphics[width=10cm]{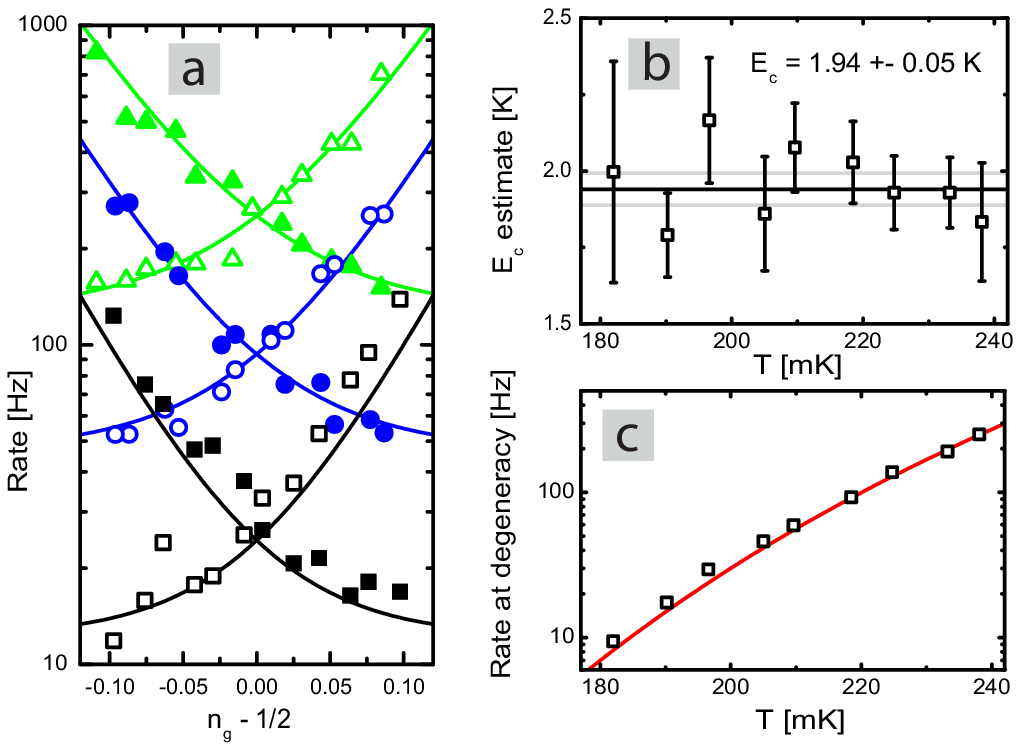}
    \caption{Observed rate of thermally activated tunneling as a function of the gate charge offset from degeneracy at bath temperatures 197, 218, and 238 mK. Open and filled symbols correspond to $0 \rightarrow 1$ and $1 \rightarrow 0$ transitions, respectively. Solid lines show the theoretical prediction according to Eq.~(\ref{eq:GammaE}) with fitted sample parameters. (b) Estimates for the box charging energy $\Ec$ obtained by application of the detailed balance to the observed $\Gamma_{0\rightarrow 1}/\Gamma_{1\rightarrow0}$ ratio as a function of gate offset at different temperatures. (c) Observed tunneling rate at degeneracy as a function of temperature (square markers), and best fit to Eq.~(\ref{eq:GammaE}) (solid line).}
\end{figure}

\begin{figure}
    \includegraphics[width=10cm]{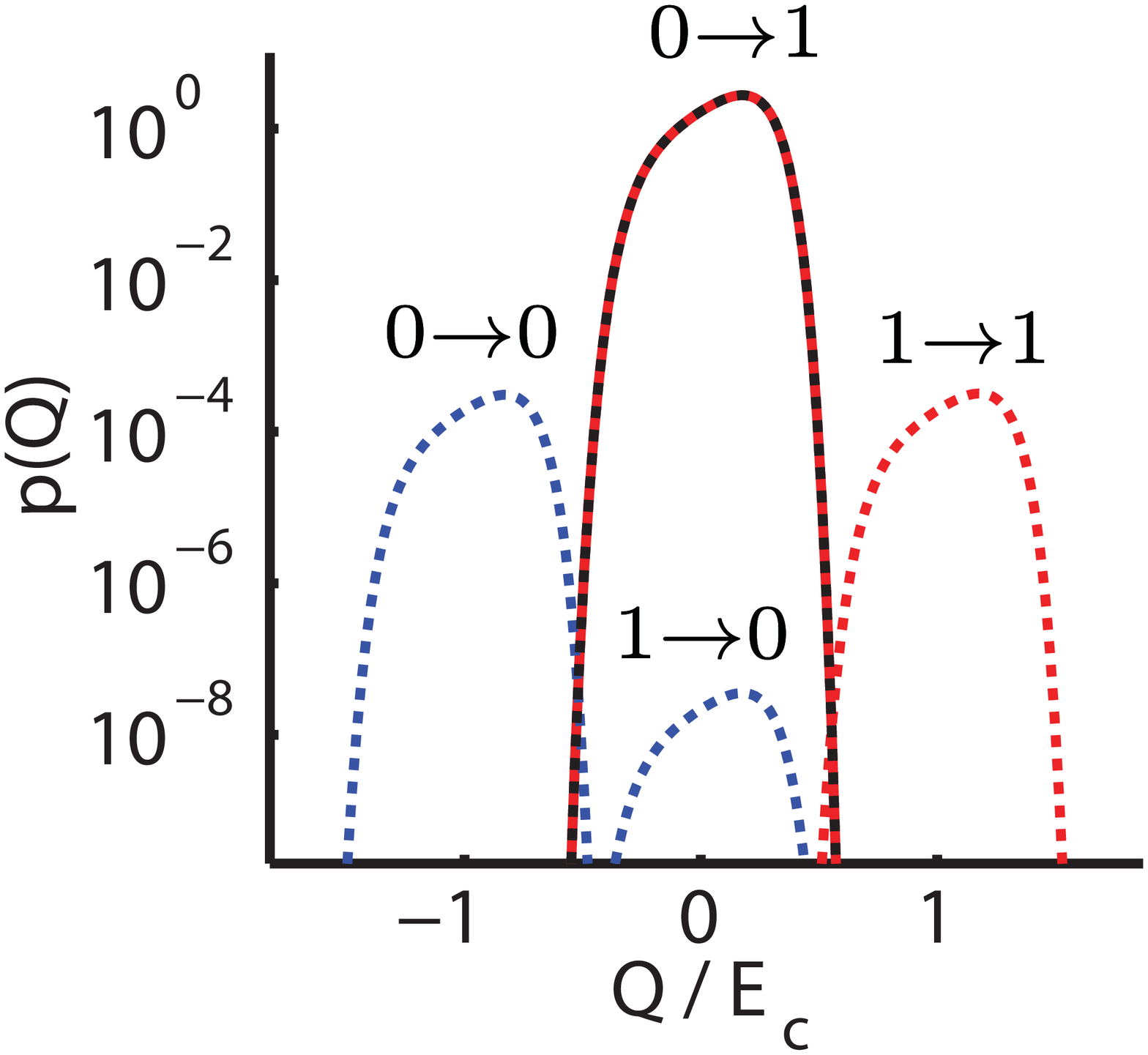}
    \caption{Full simulated $Q$ distributions for the four different trajectory types weighted with their occurrence probability plotted as dashed lines. The simulations were made for a $f = 5~\mathrm{Hz}$ sinusoidal drive at $T =$ 214~mK, and the sample parameters and operation conditions correspond to those stated in the main article. In Fig.~2(a) of the main article, only the $Q_{0\rightarrow1}$ distributions are presented. Here, the full distribution of $W - \Delta F$ is overlaid on the graph with a solid line [with $(W - \Delta F)/\Ec$ as the x axis], and it is indistinguishable from that of $Q_{0\rightarrow1}$.}
\end{figure}